\documentstyle[sprocl]{article}
\input epsf

\def\be{\begin{equation}}
\def\ee{\end{equation}}
\def\bea{\begin{eqnarray}}
\def\eea{\end{eqnarray}}

\begin{document}

\title{SIGNATURES of DYNAMICAL SYMMETRY BREAKING at TEVATRON
UPGRADE and LHC}

\author{ S. DE CURTIS}

\address{Istituto Nazionale di Fisica Nucleare, Sezione di Firenze,
\\ I-50125 Firenze, Italy}

\maketitle\abstracts{
We study the phenomenology of a strongly interacting electroweak symmetry breaking
sector in which the physics is dominated by spin one resonances. 
Specifically we will consider an effective description of dynamical symmetry breaking
based on a particular model which passes all the low-energy precision tests
and gives clear signals at LHC but also at Tevatron Upgrade.}

\section{Introduction}

\def\lq{\left [}
\def\rq{\right ]}
\def\gp{g'}
\def\gs{g''}
\def\rs{\sqrt{s}}
\def\eps{\epsilon}
\def\ggs{\frac{g}{\gs}}
\def\f{\frac}
\def\lmu{L_{\mu}}
\def\rmu{R_{\mu}}
\def\ct{c_\theta}
\def\st{s_\theta}
\newcommand{\nn}{\nonumber}
\newcommand{\dd}{\displaystyle}
\font\twelve=cmbx10 at 15pt
\font\ten=cmbx10 at 12pt
\font\eight=cmr8
\def\build#1_#2^#3{\mathrel{\mathop{\kern 0pt#1}\limits_{#2}^{#3}}}
\def\vect#1{\overrightarrow{#1\kern 1pt}\kern-1pt}
\newcommand{\rf}[1]{(\ref{#1})}

\def\wt{\widetilde}
\def\wh{\widehat}
\def\ul{\underline}
\def\ol{\overline}
\def\tg{\mathop{\rm tg}\nolimits}
\def\cotg{\mathop{\rm cotg}\nolimits}
\arraycolsep2pt

We present some results on the usefulness of upgraded Tevatron and LHC
hadron colliders 
to test the idea of a strongly interacting sector as responsible for
the electroweak symmetry breaking.

The calculations are performed within an effective
lagrangian description, called the BESS model~\cite{bess}
(BESS standing for Breaking
Electroweak Symmetry Strongly), which provides for a
rather general frame based on the
standing point of custodial symmetry and gauge invariance, without
specifying any dynamical scheme.
We are interested in studying a spontaneous symmetry breaking
avoiding physical scalar particles. 
An effective lagrangian describing in an unified way
mass terms and interactions of the standard electroweak gauge bosons
can be derived as a gauged non linear
$\sigma$-model.
Using extensively the fact that any non linear
$\sigma$-model is gauge equivalent to theories with additional
hidden local symmetry~\cite{bando}, we introduce new vector
resonances, similar to ordinary $\rho$ vector mesons or to
the techni-$\rho$ particle of technicolor theories, as the
gauge bosons associated to the hidden symmetry group of $SU(2)$
type. Under the assumption they are dynamical, we will get the 
$SU(2)$ minimal BESS model~\cite{bess} 
described by a Yang Mills lagrangian whose gauge group is $SU(2)_L
\otimes U(1)_Y \otimes SU(2)_V$.
 The  parameters it contains, besides the standard model (hereafter denoted as SM)
ones,  are
the mass $M_V$ of the new bosons forming a degenerate $SU(2)$
triplet and
their gauge coupling constant $g''$.
The new particles are naturally coupled to fermions through mixing
between $ W$, $Y$ and $V$, although a direct
coupling, specified by a new parameter $b$, is possible.
The SM
  is recovered in the limit $g'' \to \infty$ and
$b = 0$. Mixings of the ordinary gauge bosons to the $V$'s are 
  ${\cal O}\left( {g}/{g''} \right)$. Due to these mixings, the $V$ bosons are
coupled to fermions even for $b = 0$. Furthermore these couplings are
still present in the limit $M_V \to \infty$, and therefore the new
gauge boson effects do not decouple in the large mass limit.

The model can also incorporate axial-vector
resonances by enlarging the additional gauge group from
$SU(2)_V$ to $SU(2)_L \otimes SU(2)_R$ local~\cite{assiali}.
 Also in this case the Goldstone
bosons associated to the breaking of the gauge group to the $U(1)_{em}$
are absorbed by the vector and the axial-vector
bosons and by the standard $W$ and $Z$.
In this extension of the model we have two additional parameters:
 the mass $M_A$ of
the axial-vector resonances and $z$ the ratio of the mixings
of $V$ and $A$ bosons to the $W$.

The detailed study of the symmetries of the effective theory shows however 
that in special cases the resulting symmetry can be larger than the one 
requested by the construction. For the particular choice $M_V=M_A$ and $z=1$
 a maximal symmetry $[SU(2)\otimes SU(2)]^3$ is realized for
the low energy effective lagrangian~\cite{dege}.
This case is called degenerate BESS model and it        
turns out to be very useful 
in relation to schemes of strong electroweak breaking.

 We stress
immediately its main property and what makes it so attractive: in degenerate
BESS all deviations in the low 
energy parameters from their standard model  values are strongly 
suppressed. This would make it possible that a strong electroweak sector at 
relatively low energies exists within the precision of electroweak tests, 
such that it may be accessible with existing accelerators (Tevatron)
or with accelerators
projected for the near future (LHC). In fact one can show that the lagrangian of
degenerate BESS becomes identical to that of the standard model
(except for the Higgs sector) for
sufficiently large mass of the degenerate vector and axial-vector mesons. In
other words, different from the minimal BESS ~\cite{bess}, where
such a high mass decoupling is not satisfied, the decoupling occurs in
degenerate BESS. 

The phenomenological implications of the degenerate BESS will be 
a substantial part of our discussion below.

\section{Degenerate BESS model}

In general, the
existence of new vector and axial-vector bosons 
 indirectly manifests at LEP through deviations from SM
expectations~\cite{self}. For this purpose a low energy effective theory
valid for heavy resonances~\cite{anich} is useful.

As well known, in the low energy limit, one can parameterize the modifications
due to a heavy sector in terms of three independent parameters: $\Delta r_W$,
$\Delta k$, $\Delta \rho$, or equivalently
$\eps_1$, $\eps_2$, $\eps_3$~\cite{alta}. 
In the minimal BESS model neglecting terms ${\cal O}
(M_Z^2/M_V^2)$) one gets: $\eps_1 = \eps_2 = 0$
while 
$\eps_3 = (g/ \gs)^2 - b/2 $
is the sum of two contributions: one given by the mixing and the other by the direct
coupling of $V$ bosons to fermions. By comparing with the experimental value of the 
$\eps_3$ parameter as obtained by a global fit to all the available experimental data 
(expecially from LEP) we find that
 the present bounds on the minimal BESS model are
quite stringent~\cite{anich} and, unless one considers a sort of fine 
tuning between the
two parameters, very small values of $b$ and $g/\gs$ are still allowed
(of the order of few per cent). 
A question is natural:
is it possible to think of a model af strongly interacting symmetry breaking 
sector which avoids the restrictive bounds from LEP?

We recall that the very small experimental value of the $\epsilon_3$
parameter, which measures the amount of isospin-conserving virtual
contributions to the vector boson self-energies, strongly
disadvantages the ordinary technicolor schemes, for which the contribution to
$\epsilon_3$ is large and positive.
This problem could be attributed to the vector dominance
in the dispersion relation satisfied by the $\eps_3$ parameter~\cite{S}:
\be
\eps_3=-\frac{g^2}{4 \pi} \int_0^\infty \frac{ds}{s^2}
[Im \Pi_{VV}(s)- Im \Pi_{AA}(s)]
\ee
where $\Pi_{VV(AA)}$ is the correlator between two vector (axial-vector)
currents.
If the vector and axial-vector spectral functions are saturated
by lowest lying vector and axial-vector resonances, one has
$Im \Pi_{VV(AA)}(s)=-\pi g_{V(A)}^2 \delta(s-M^2_{V(A)})$
where $g_{V(A)}$ parameterizes the matrix element of the vector
(axial-vector) current between the vacuum and the state $V_\mu$ ($A_\mu$),
and $M_{V(A)}$ is the vector (axial-vector) mass.
From the previous equation, one obtains
\be
\eps_3=\frac{g^2}{4}\left[\frac{g_V^2}{M_V^4}-\frac{g_A^2}{M_A^4}\right]
\ee
By evaluating eq. (2) within the BESS model with also
axial-vector resonances~\cite{assiali} (with $b=0$)
one gets $\eps_3=(g/\gs)^2 (1-z^2)$.
If the underlying theory mimics the QCD behaviour, naively scaled
from $f_\pi\simeq 93~MeV$ to $v\simeq 246~ GeV$, then
$z=1/2$ and so  the deviation for the $\epsilon_3$ parameter is positive and
potentially large~ \cite{alta2}.
On the contrary, in the  degenerate BESS model~\cite{dege}
the approximate 
degeneracy among the masses of the vector and axial-vector 
states and their couplings makes $\eps_3$ vanish (in the low energy limit).
In other words, by calling $M$ the common mass of the resonances
(they are degenerate up to weak corrections)
in the $M\to\infty$ limit, the model decouples
and all the $\eps_i$ go to zero. 

The important feature of the model under examination is that
the previous results are protected by an  extended vector-axial symmetry
$[SU(2)\otimes SU(2)]^3$, broken at low-energy by the 
electroweak interactions (in fact $\eps_3=0$ follows 
from the $SU(2)_L\otimes SU(2)_R$ custodial symmetry \cite{inami}).

In an expansion in $M_Z^2/M^2$, we have evaluated
the leading terms for the corrections to the
$\epsilon$ parameters~\cite {dege}.
They  are  given by
\be
\epsilon_1=-\f{\ct^4+\st^4}{\ct^2}~ X,~~~~~
\epsilon_2=-\ct^2~ X,~~~~~
\epsilon_3=-X
\ee
all proportional to $X\approx 2(M_Z^2/M^2)(g/\gs)^2$
which contains a double suppression
factor: $M^2_Z/M^2$ and $(g/\gs)^2$.
Radiative corrections have also to be taken into account. We assume for the BESS 
model the same one-loop radiative corrections as for the SM in which the Higgs
mass is used as a cut-off $\Lambda$. For a top mass value of 175 $GeV$ and for 
$M_H=\Lambda=1~TeV$, we can compare the sum of the SM contributions
and the previous deviations with the experimental
 values for the $\epsilon$ parameters, 
determined from the available LEP data and the $M_W$ measurement from Tevatron:
\cite{cara}
$\epsilon_1=(3.48\pm 1.49)\cdot 10^{-3}$,
$\epsilon_2=(-5.7\pm 4.19)\cdot 10^{-3}$,
$\epsilon_3=(3.25\pm 1.40)\cdot 10^{-3}$.
From the combinations of these 
  experimental results, the upper limit in the plane $(M, g/\gs)$ is
given in Fig. 1 (solid line).
The statistical significance of the plot is that of a 95\% C.L. limit in one
variable, the mass, at a given value of $g/g''$.
The result of this analysis shows that in the degenerate BESS 
relatively light resonances are compatible with the electroweak data as given by
LEP and Tevatron.

Besides studying the virtual effects we shall also discuss the direct production
of the heavy resonances. Data from the Fermilab Tevatron Collider, 
collected by the CDF collaboration
\cite{cdf} establish limits on the model parameter space. Their search was 
done
through the decay $W' \to e \nu$, assuming standard couplings of the $W'$ to
the fermions. Their result can be easily translated into a limit for the 
degenerate
BESS model parameter space. 
In Fig. 1 these limits are
shown (dashed line). 
 The excluded region is above the curve.
The figure was obtained using the CDF 95\% C.L. limit
on the $W'$ cross-section times the branching ratio at $\sqrt{s}=1.8~TeV$
and an integrated luminosity of $19.7~pb^{-1}$, and comparing this limit 
with the predictions of our model at fixed $g/g''$, thus 
giving a limit for  $M$. This procedure 
was then iterated for various values of $g/g''$. 
 The limit from CDF is more
restrictive for low resonance masses, while LEP limit is more restrictive for
higher mass values. 
\begin{figure} 
\epsfxsize=6truecm.
\centerline{\epsffile[65 262 500 678]{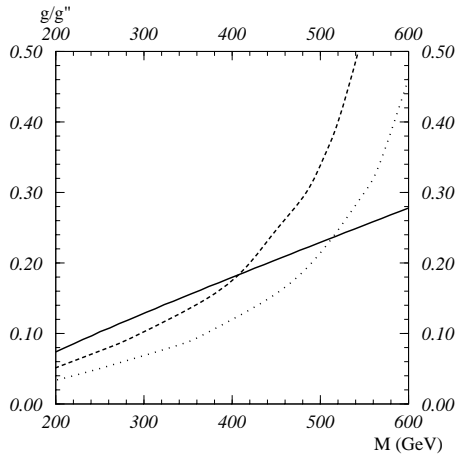}}
\caption{ $ 95 \% $ C.L. upper bounds on $g/\gs$ vs. M from 
LEP data (solid line)
and CDF with $L=19.7~pb^{-1}$ (dashed line).
The dotted line shows the extrapolation of the CDF bounds to $L=100~ pb^{-1}$.}
\end{figure}
Since in the running period just completed at Tevatron the integrated luminosity was
 more than 100 $pb^{-1}$, by waiting for the analysis of these data, we have
considered an extrapolation of the CDF limit in the electron channel.
This extrapolation is based on the simple principle that when background is present
the cross-section limit scale inversely with the square root of 
the luminosity~\cite{TeV2000}.
The result is given by the dotted line in Fig. 1.  

We have also studied the sensitivity of degenerate BESS at LEP2
by comparing cross-sections and asymmetries in the 
fermionic pair channels and $WW$ channel between the model and SM. 
The general conclusion~\cite{dege} is that the bounds on the model would not be much 
stronger than those from LEP and Tevatron. 

We conclude this section with some remarks about the decay
of the vector mesons $\lmu$ and $\rmu$ ($L_\mu=(V_\mu-A_\mu)/2$, $R_\mu=(V_\mu+A_\mu)/2$).
 A feature of degenerate BESS, as compared to BESS with only vector resonances, 
comes from the absence of direct coupling of the new resonances to the 
 would-be Goldstone bosons which provide the longitudinal
degrees of freedom to the $W$'s, then their partial width 
into longitudinal $W$'s will be suppressed. As a consequence
the width into a $W$ pair is of the same order of the fermionic width.
Unlikely other schemes of strong
electroweak breaking (as the minimal BESS model), the $W_L W_L$ channel is not
enhanced. On the contrary we expect very good signatures from the degenerate BESS
in the di-lepton channels.

\section{Degenerate BESS at Tevatron Upgrade and LHC}

We have considered the detection of a signal from strong
electroweak sector at a possible upgrading of the Fermilab
Tevatron~\cite{TeV2000}. The option we have chosen, is the so called TeV-33 with 
a c.m. energy of the collider of  2 $TeV$ and 
an integrated luminosity of $10~fb^{-1}$.
We have analyzed the production of the charged resonances $L^\pm$ of the degenerate
 BESS in the leptonic channel.
The choice of  this channel is due to the 
clean signature and the large cross-section.
The events where simulated using 
Pythia Montecarlo \cite{phy}. A rough simulation of the detector was 
also performed. The energy of the leptons was smeared according to
$\Delta E/E = 15\%$
and the error in the 3-momentum determination was assumed of 5\%.
Only Drell-Yan mechanism for production was considered and the signal events are
compared with the background from SM di-lepton production.
For example, in Fig. 2 we show the  
 differential distribution of events of 
$p\bar p \to L^\pm, 
W^\pm \to
\mu \nu$  in the transverse momentum of the muon for $M_{L\pm}=500~GeV$
and $g/\gs=0.2$. Even if the cuts applied are not
optimized (see Table 1), the Jacobian peak is very well 
visible over the SM background.
\begin{figure} 
\epsfxsize=6truecm
\centerline{\epsffile[60 200 540 650]{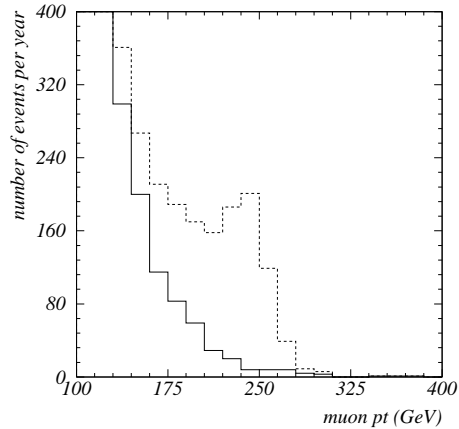}}
\caption{ ${(p_t)}_{\mu}$ distribution of $p {\bar p}\to L^\pm,W^\pm\to \mu\nu$
events at Tevatron Upgrade for $M_{L^\pm}=500~GeV$ and $g/\gs=0.2$.
The following cuts have been applied: $\vert p_{t\mu}\vert,
\vert p_{t~miss}\vert >100~GeV$, $E^{miss}>50~GeV$.
The solid line is the SM background, the dashed line is the degenerate BESS
model signal.}
\end{figure} 

We have examined various cases with different choices of $M$, and $g/g''$
(taken inside the physical region shown in Fig. 1)
to give an estimate of the sensitivity of the model to this
option for the upgrading  of the Tevatron (see Table 1).
For each case we have applied optimal cuts to maximize the statistical significance 
of the signal (last column in Table 1).
We see that the number of signal events decreases for increasing mass of the 
resonance. The conclusion is that Tevatron with the high luminosity option could be 
enough to discover a strong electroweak resonant sector as described by the
degenerate BESS model for masses up to 1 $TeV$. 
\begin{table}[t] 
\caption{Degenerate BESS at Tevatron Upgrade. For all the cases
we have also applied a cut $\vert p_t^{miss}\vert_c=\vert p_t^\mu\vert_c$.
 Here $\# B(S)$=number of background(signal) events. }
\begin{center}
\begin{tabular}{|c|c|c||c|c||c|c|c|}
\hline 
&&&&&&&\\
$\dd{\f{g}{g''}}$ & $M_{L^\pm}$ & $\Gamma_{L^\pm}$ & 
$\vert p_t^\mu\vert_c$ & 
$(E^{miss})_c$ &  $\# B$ & $\# S$ & $\dd{\f{S}{\sqrt{S+B}}}$ \\
&(GeV)&(GeV)&(GeV)&(GeV)&&&\\ 
\hline 
\hline 
0.12 & 400 & 0.4 & 100 & 150 & 1271 & 1135 & 23.1\\  
\hline
0.20 & 500 & 1.4 & 150 & 200 & 223 & 870 & 26.3\\  
\hline
0.20 & 600 & 1.6 & 200 & 100 & 69 & 282 & 15.0\\  
\hline
0.30 & 800 & 4.9 & 250 & 100 & 19 & 83 & 8.2\\  
\hline
0.40 & 1000 & 10.9 & 250 & 100 & 17 & 21 & 3.4\\  
\hline
\end{tabular}
\end{center}
\end{table}

The same analysis performed at the LHC collider 
gives very clear signatures.
We have considered a configuration of LHC with a c.m. energy
$\sqrt{s}=14~TeV$,
a luminosity of $10^{34} cm^{-2} s^{-1}$  and one year run ($10^7$ s). 
 Such a
machine will be able either to discover the new resonances or to constrain
the physical region left unconstrained by previous data.
\begin{figure} 
\epsfysize=6truecm
\centerline{\epsffile[50 200 538 653]{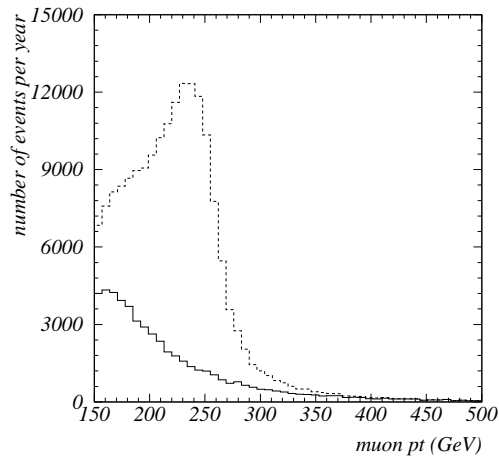}}
\caption{${(p_t)}_\mu$ distribution of $pp \to L^\pm, W^\pm \to
\mu \nu$ events at LHC, for $M_{L^{\pm}}=500~GeV$, $g/g''=0.15$.
The following cuts have been applied: $\vert p_{t\mu}\vert ,
\vert p_{t~miss}\vert >150~GeV$. The continuous line represents the 
Standard Model background while the dashed one is the degenerate BESS model 
expectation.}
\end{figure}
In Fig. 3
we show the differential distribution of events at LHC of 
$pp \to L^\pm, 
W^\pm \to
\mu \nu$  in the transverse momentum of the muon for 
 a spectacular case corresponding to 
$M_{L^{\pm}}=500~GeV$ and $g/g''=0.15$. The total $L^{\pm}$ width is 
$\Gamma_{L^{\pm}}=0.9~GeV$, with the corresponding 
$B(L^+\to\mu\nu)=8.5\times 
10^{-2}$.  The number of signal events per year 
is approximately 128000, the corresponding background consists of 51500 
events.
Also for higher masses of the resonances we have a very large number of events 
in the LHC configuration~\cite{dege}.
Clearly these results are a little bit optimistic. In fact
 the reconstruction of resonance mass will require a careful study 
of 
the experimental setup, due to the smallness of the resonance width. Anyway
we think that since the signal to background  ratio is extremely good 
there would not be problems for detectability.

Finally, we have 
considered the possibility that no new resonances are discovered at LHC.
In this case limits can be imposed on the parameter space of the model. 
We have calculated the total cross-section $pp \to L^\pm, W^\pm \to
\ell \nu$ and, and compared with the SM background~\cite{dege}. The result is that
the new resonances of the model can be discovered directly for a wide range 
of values of the parameter space of the model. The discovery limit in the 
mass of the resonance depends on the value of $g/g''$. 
For example if $g/g''=0.1$, the 
resonance is visible over the background at least up to 2 $TeV$, in the 
channel $pp \to \mu \nu$.

In this preliminary study we did not consider the production and decay 
of the corresponding neutral resonances of the model. Another subject
under study is the hadronic decay channel which could give much more informations
expecially at Tevatron Upgrade.  

\section{Conclusions}

We have used the BESS model, as a rather general frame based on custodial
symmetry and gauge invariance, to examine the possibilities offered by 
the  Tevatron Upgrade and by LHC, 
 to test for strong electroweak breaking.
In particular the calculations are performed within the degenerate BESS,
a model which, due to its decoupling properties, passes all the low energy 
precision tests. 
We have studied the direct production at hadron colliders
of the new charged resonances decaying 
leptonically. By comparing with the SM background,
 within suitable cuts and detection limits we conclude that Tevatron with the
high luminosity option could be enough to discover a strong electroweak resonant 
sector as described by the degenerate BESS for masses up to 1 $TeV$.
The same analysis performed at LHC gives very spectacular signatures. On the other
hand, if no resonances will be discovered at LHC the parameter region of
the degenerate BESS will practically close.

The degenerate BESS model is comparatively much more
evident than ordinary BESS, and probably than any other strong electroweak 
model not sharing its peculiar symmetry properties.

\section*{Acknowledgments}
I would like to thank 
A. Deandrea, R.Casalbuoni, D.Dominici and R.Gatto for enlightening discussions
and, in particular, A.Deandrea for the help in producing the figures.

\end{document}